\documentclass[twocolumn,prl,aps,amsmath,amssymb,nofootinbib]{revtex4}
\usepackage{graphicx}
\usepackage{dcolumn}
\usepackage{bm}
\usepackage{mathrsfs}
\usepackage{slashed}
\usepackage{amsmath,lscape,epsfig}
\usepackage{amsfonts}
\usepackage{amsmath}
\usepackage{amssymb}

\begin{document}

\title{Analog model for quantum gravity effects: phonons in random fluids}
\author{
G. Krein$^1$\footnote{gkrein@if.ufrj.br},
G. Menezes$^1$\footnote{gsm@ift.unesp.br}
and N.F. Svaiter$^2$\footnote{nfuxsvai@cbpf.br}}
\affiliation{$^1$Instituto de F\'\i sica Te\'orica, Universidade Estadual Paulista\\
Rua Dr. Bento Teobaldo Ferraz 271 - Bloco II, 01140-070  S\~ao Paulo, SP , Brazil \\
$^2$Centro Brasileiro de Pesquisas F\'{\i}sicas, Rua Dr. Xavier Sigaud 150,
22290-180  Rio de Janeiro, RJ, Brazil }

\begin{abstract}
We describe an analog model for quantum gravity effects in condensed
matter physics. The situation discussed is that of phonons
propagating in a fluid with a random velocity wave equation. We
consider that there are random fluctuations in the reciprocal
of the bulk modulus of the system and study free phonons in the
presence of Gaussian colored noise with zero mean. We show
that in this model, after performing the random averages over
the noise function a free conventional scalar quantum field theory
describing free phonons becomes a self-interacting model.
\end{abstract}

\pacs{04.60.-m, 04.62.+v, 42.25.Dd}

\maketitle

Unruh has shown that the propagation of sound waves in a
hypersonic fluid is equivalent to the propagation of scalar waves
in a black hole spacetime~\cite{unruh1}. Quantizing the acoustic
wave in such physical system with sonic horizon implies that the
sonic black hole can emit sound waves with a thermal spectrum,
recovering the Hawking result \cite{hawking} in this analog model,
i.e. the presence of phononic Hawking radiation from the acoustic
horizon. Since the Unruh fundamental paper, the possibility of
simulating aspects of general relativity and quantum fields in
curved spacetime through analog models has been widely discussed
in the literature \cite{mario,Barcelo:2005fc}. In the context of
quantum gravity, most of the work on
analogue models has focussed attention on
violation of Lorentz invariance. The key point of the
discussion is the expectation of a low-energy
manifestation of a possible spacetime discreteness at the Plank
scale. Such a manifestation is precisely violation of Lorentz
invariance. In these analog models interesting dispersion
relations are derived~\cite{arte,qg1}.

The aim of the present paper is to take a somewhat different look at
the way an analog model for quantum gravity effects can be
implemented. Here we have in mind the results obtained by Ford and
collaborators \cite{ford11}, where fluctuations of the geometry of
spacetime caused by a bath of squeezed gravitons has the effect of
smearing out the light cone. In addition to quantum mechanical
metric fluctuations, there are induced metric fluctuations generated
by quantum fluctuations of matter fields. In the regime where
induced metric fluctuations dominate, the conventional approach is
to assume a stochastic ensemble of fluctuating geometries. In this
framework, Hu and Shiokawa \cite{hu1} assumed that the metric tensor
has a deterministic and a stochastic contribution \cite{principal}.
We argue that in condensed matter physics there is an analog model
for fluctuations on the light cone generated by a thermal bath of
squeezed gravitons. We consider a disordered medium with random
classical fluctuations in the reciprocal of the bulk modulus.
Actually, the analog model we discuss in the paper is closely
related to the induced metric fluctuations framework. We consider a
scalar quantum field theory associated to acoustic waves defined in
a fluid with classical random fluctuations, where the velocity of
propagation of the acoustic waves fluctuates randomly, and
consequently the sonic cone fluctuates.

Fluctuations of a fluid consist of two types, classical and quantum.
Quantum fluctuations are due to quantization of fluid thermodynamic
observables. Consideration of quantum fluctuations motivated Ford
and Svaiter~\cite{ford1} to discuss the scattering of light by the
zero-point density fluctuations of a phonon field. Also, since there
is an analog to the Casimir effect when the zero-point fluctuations
of the phonon field are submitted to boundary conditions
\cite{lifshitz}, the same authors considered in Ref.~\cite{ford2}
the effects of boundaries discussing the density correlation
function associated to the phonon field in many different
situations. In the present paper we will show that fluids are not
only useful analog models for effects in field theory, but random
fluids with particular properties can simulate quantum gravity
effects.

Let us start from the basic equations of fluid dynamics \cite
{fluid}. The local mass density $\rho(t,\vec{r})$, the pressure
$p(t,\vec{r})$, the local velocity $\vec{v}(t,\vec{r})$ and the
local temperature $T(t,\vec{r})$ are the thermodynamic fields of
the system. For an ideal fluid, assuming thermodynamical
equilibrium, the acoustic wave equations are obtained by
linearizing the fluid dynamics equations for small disturbances
around the fluid at rest, namely
\begin{eqnarray}
p(t,\vec{r}) &=& p_{0}+\delta p(t,\vec{r}), \label{p8}
\\
\rho(t,\vec{r}) &=& \rho_{0}+\delta\rho(t,\vec{r}), \label{p9}
\\
\vec{v}(t,\vec{r}) &=& \delta\vec{v}(t,\vec{r}),
\label{p10}
\end{eqnarray}
where $\rho_{0}$ and $p_{0}$ are the constant equilibrium density
and pressure. Using the Euler and mass balance equations, one obtains
\begin{eqnarray}
&& \frac{\partial}{\partial t}
\delta\vec{v}(t,\vec{r}))+\frac{1}{\rho_{0}}{\nabla}\delta
p(t,\vec{r})=0,
\label{p11}
\\
&&
\frac{\partial}{\partial t}\,\delta
\rho(t,\vec{r})+\rho_{0}{\nabla} \cdot \delta\vec{v}(t,\vec{r})=0.
\label{p12}
\end{eqnarray}
The bulk modulus $K$ of the fluid is defined as
$K = \rho_0 {\partial p}/{\partial \rho}= \rho_0 u^2$,
where $u=({\partial p}/{\partial \rho_{0}})^{1/2}$ is the sound
velocity. Using that the sound wave in a ideal fluid is adiabatic,
${\delta p(t,\vec{r})}/{\delta \rho(t,\vec{r})} =
{\partial p(t,\vec{r})}/{\partial \rho_{0}}$, one obtains from
the fluid equations a wave
equation for $\psi(t,\vec r) \equiv \delta p(t,\vec r)$
\begin{equation}
\left(\frac{1}{u^2}\frac{\partial^2 }{\partial t^2} - \nabla^2 \right)
\psi(t,\vec r) = 0.
\label{eq-psi}
\end{equation}
Here, in general $u$ can depend on $\vec r$. This wave equation, as
can be easily checked, is similar as the wave equation one would
obtain if considering a curved spacetime where the metric contains a
deterministic and a stochastic contribution, similar to the
situation discussed in Ref.~\cite{hu1}, where metric fluctuations
were represented as fluctuations in the optical index associated to
electromagnetic waves. This equation also appears in the study of
elementary excitations in disordered elastic media with random
fluctuations, which has been discussed recently by many authors. For
example, it has been pointed out that a free phonon field in a
disordered elastic medium becomes self-interacting~\cite{gurarie}.
As we will show in the following, a similar situation occurs when
$u(\vec r)$ fluctuates.

When $u$ is a constant, one can derive a wave equation also for
$\delta \rho(t,\vec r)$ and $\delta \phi(t,\vec{r})$, with
$\delta\vec v = \nabla \delta\phi$. In this case the quantization of
the system follows as usual~\cite{gorkov}, the phonons have a linear
dispersion relation $\omega(\vec k) = u \, |\vec{k}|$ and the
operators $\delta \rho(t,\vec r)$ and $\delta \phi(t,\vec r)$ obey
equal-time commutation relation given by [we assume $\hbar=1$]
\begin{equation}
[\,\delta\phi(t,\vec{r}),\delta\rho(t,\vec{r}\,')\,]
=-i\delta^{3}(\vec{r}-\vec{r}\,').
 \label{p17}
\end{equation}
The unperturbed two-point Green's function of the phonon field
$D_F(x,x')$ is given by the expectation value of the time ordered
product
\begin{equation}
D_F(x,x')=-i\langle
0|T(\delta\rho(t,\vec{r})\delta\rho(t',\vec{r}\,'))|0\rangle,
\label{p22}
\end{equation}
where $|0\rangle$ is the ground state of the phonon field.

Having performed a linearization to obtain a wave equation for the
acoustic perturbation, let us consider the influence of random
fluctuations over the phonon field. Under such circumstances, the
fluid equations become random differential equations. The simplest
assumption one can make is to consider a quantized phonon field in a
fluid with bulk modulus that fluctuates randomly, with $\rho_0$ kept
constant \cite{papa}. This generates fluctuations in the velocity of
the acoustic perturbations. Therefore, let us assume (for simplicity
we shall consider one-dimensional fluids, the extension to three
dimensions is not difficult)
\begin{equation}
\frac{1}{u^2(z)}= \frac{1}{u_{0}^2}\bigl(1+\nu(z)\bigr),
\label{p28}
\end{equation}
with $u_0$ the sound speed in a homogenous medium and $\nu(z)$ is
taken to be a Gaussian colored noise with zero mean
\begin{equation}
\langle \nu(z) \rangle_{\nu} = 0, \hspace{0.5cm}
\langle \nu(z)\nu(z') \rangle_{\nu} = \sigma^2 \, h_{\Lambda}(z-z').
\label{p29}
\end{equation}
The symbol $\langle\, ...\,\rangle_{\nu}$ denotes average over
noise realizations. Under such circumstance, the fluid equations
become random differential equations and two important points
should be noted. First, one needs to specify a stochastic modeling
of the fluctuations of the medium parameters in terms of random
processes. Second, the scale separation in the problem should be
properly addressed. This second point will be discussed latter
ahead in the paper. With respect to the first point, we mention
that the specific form of $h_{\Lambda}$ is not relevant for our
arguments, it is sufficient to note that $\Lambda \sim 1/l_d$,
where $l_d$ is the correlation length of the noise field, and for
$\Lambda \rightarrow \infty$ one has uncorrelated white noise.
However, it should be clear that $\nu(z)$ is modeling fluctuations
of a spacetime background on which a quantum field is propagating
and it should be related to geometry fluctuations on a more fundamental
level. For example, two-dimensional geometries are characterized by the
scalar curvature $R$ of the Riemann tensor and one could speculate
that $\nu(z)$ should be related to $R$. We will not try to make
such a connection here because this would take us away from the
main scope of the paper.

Our aim is to calculate the density-density correlation function,
which is related to an observable as e.g. in a light (or neutron)
scattering experiment. This is most easily calculated solving first
Eq.~(\ref{eq-psi}) and then using the relation $\psi(t,z) = u^2(z)\,
\delta \rho(t,z)$. We solve Eq.~(\ref{eq-psi}) by a conventional
perturbation expansion in $\nu(z)$. We write $\psi(t,z) =
\sum_{n=0}^{\infty}\,\psi_{n}(t,z)$,
where $\psi_n(t,z)$ is of order $n$ in~$\nu(z)$. Substituting
this in Eq.~(\ref{eq-psi}) and collecting terms of the same
order in $\nu(z)$, one obtains the recursive relation for $n > 0$
\begin{equation}
\Biggl(\frac{1}{u_0^{2}}\frac{\partial^{2}}{\partial
t^{2}}-\frac{\partial^{2}}{\partial z^2}\Biggr)\psi_n(t,z) =
-\frac{\nu(z)}{u_0^2}\frac{\partial^2}{\partial
t^2}\,\psi_{n-1}(t,z). \label{p36}
\end{equation}
For $n = 0$, we have the standard wave equation for
$\psi_{0}(t,z)$. Choosing the solution of the homogeneous equation
to be zero, the general solution of Eq.~(\ref{p36}) can be written
as
\begin{equation}
\psi_n(t,z) =
\int dz' dt'\,G_F(t-t';z-z')\frac{\nu(z')}{u_0^2}\frac{\partial^2}{\partial
t'^2}\psi_{n-1}(t',z'), \label{p37}
\end{equation}
where $G_F(t-t';z-z')$ is the Green's function associated with the
differential operator on the l.h.s. of Eq.~(\ref{p36}). Introducing
a Fourier representation for $\psi_n(t,z)$
\begin{equation}
\psi_n(t,z) = \int \frac{d\omega}{2\pi} \,
e^{-i\omega t}\,\psi_n(\omega,z), \label{p39}
\end{equation}
and a similar one for $G_F(t-t';z-z')$, the complete solution
can be written as
\begin{widetext}
\begin{eqnarray}
\psi(\omega,z) = \psi_0(\omega,z) +
\,\sum_{n=1}^{\infty}(-1)^{n}\bigg(\frac{\omega}{u_0}\bigg)^{2n}
\int\,\prod_{m=1}^n\,dz_m\,G_F(\omega;z_{m-1} -
z_m)\nu(z_m)\,\psi_0(\omega,z_n), \label{p42}
\end{eqnarray}
\end{widetext}
where it is to be understood that $z_0 = z$.

Given the solution, one can calculate the density-density
correlation function using $\delta \rho(t,z) = \psi(t,z)/u^2(z)$ and
quantize as usual. It is important to call attention to the validity
of the linear dispersion relation in the case of randomness. This
situation is similar to the relation of the light cone fluctuations
to Lorentz symmetry. In the same way as light cone fluctuations
respect Lorentz symmetry on the average, the linear dispersion
relation leads to $k^{2}/\omega^{2}=(1+\nu(z))\,
k_0^2/\omega_{0}^2$, which should be considered as valid on the
average.

Before proceeding, let us abstract for the moment from the
fact that $\psi$ represents the pressure field of the fluid and consider
it as a scalar field {\em per se}. Eq.~(\ref{eq-psi}) represents a wave
equation for a free scalar field in a fluctuating spacetime metric.
Suppose one quantizes the field $\psi$ and calculate the two-point
correlation function by taking the statistical and quantum averages
$\langle\langle\, ...\,\rangle_{\nu}\rangle$, i.e. average over noise
realizations and vacuum expectation value of products of field operators.
The two-point correlation function up to second order in $\nu(z)$ would
the be given by (in terms of the Fourier transforms of $G_F$ and
$h_\Lambda$)
\begin{equation}
\langle\langle\psi(\omega,k)\psi(\omega,k')\rangle_{\nu}\rangle =
(2\pi)^2\delta(\omega-\omega')\delta(k-k') S_{\psi}(\omega,k),
\end{equation}
where
\begin{equation}
S_{\psi}(\omega,k) = iG_F(\omega,k) + iG_F(\omega,k) \Pi(\omega,k)
iG_F(\omega,k),
\end{equation}
with
\begin{equation}
\Pi(\omega,k) = - 2 \sigma^2\,\frac{\omega^4}{u_0^4}
\int dq\, iG_F(\omega,q) h_{\Lambda}(k - q) ,
\label{p50a}
\end{equation}
where we used the fact that $\langle0|T(\psi_0(t,z)\psi_0(t',z'))|0\rangle
= i G_F(t-t',z-z')$.

It is clear that Eq.~(\ref{p50a}) leads to the interpretation that a
non-interacting scalar field propagating in a fluctuating metric
turns it into a self-interacting field. Notice that our results are
valid only if the background fluctuations are much larger than those
fluctuations defined in equations (\ref{p8})-(\ref{p10}).
Nevertheless, the scale separation in the problem mentioned above
can be analyzed in terms of its low- and high-frequency regimes. We
notice that the contributions of the perturbations to the two-point
(and $n$-point) functions have a dependence on $\omega$ in a
multiplicative way. More precisely, the deviation from a uniform
medium occurs multiplicatively with frequency. The consequences of
this observation are clear; for large frequencies, the contributions
of the perturbations dominate. In the opposite regime, for small
frequencies, the acoustic waves scatter weakly, which implies that,
for $\omega\rightarrow 0$, the field remains essentially
non-interacting. At this point, one may wonder where the correlation
length $l_{d}$ fits in this whole scenario. From Eq.~(\ref{p50a}),
we see that, for $\Lambda\rightarrow\infty$ the loop integral is
ultraviolet divergent. So, introducing correlations between the
classical random fluctuations $\nu(z)$ leads to regularized
outcomes. Therefore a physically motivated way to obtain a
non-divergent result is to use colored noise rather than a white
noise. Another point is that in principle one could think of two
different ways of performing the joint quantum and noise averages.
In a static disordered medium one could first perform the average
over quantum fluctuations for a given frozen disorder configuration
and then average over the disorder configurations~\cite{lodahl}.
This would correspond in random magnetic systems to the case of a
quenched disorder. In the fluctuating medium considered in the
present paper, the quantum and disordered averages cannot be
separated and must be performed simultaneously -- see e.g.
Ref.~\cite{skipetrov}. This case is exactly the case of annealed
systems, where the impurity degrees of freedom are in equilibrium
with the others degrees of freedom of the system.

Coming back to fluid dynamics, the density-density correlation function
can be calculated from the relation $\psi(t,z) = u^2(z) \delta \rho(t,z)$.
From Eqs.~(\ref{p28}) and (\ref{p42}), one has
\begin{widetext}
\begin{eqnarray}
\delta \rho(\omega,z) &=& \delta\rho_0(\omega,z) +
\nu(z)\delta\rho_0(\omega,z) +
\,\sum_{n=1}^{\infty}(-1)^{n}\left(\frac{\omega}{u_0}\right)^{2n}
\int\,\Biggl[\prod_{m=1}^n\,dz_m\,D_F(\omega,z - z_m)\nu(z_m) + \nonumber\\
&& + \,\nu(z) \int\,\prod_{m=1}^n\,dz_m\,D_F(\omega,z -
z_m)\nu(z_m)\Biggr]\delta\rho_0(\omega,z_n), \label{p42a}
\end{eqnarray}
\end{widetext}
where $\delta\rho_0(\omega,z) = \psi_{0}(\omega,z)/u_0^2$ and
we used $D_F$ for the $\delta\rho(t,z)$ field since its Green's
function obeys the same equation as $G_F$ for the $\psi(t,z)$
field. Taking the noise and quantum averages, the correlation
function up to second order in $\nu(z)$ is given in Fourier
space by
\begin{equation}
\langle\langle\delta\rho(\omega,k)
\delta\rho(\omega',k')\rangle_{\nu}\rangle =
(2\pi)^2\delta(\omega-\omega')\delta(k-k') S_\rho(\omega,k),
\end{equation}
with
\begin{eqnarray}
S_\rho(\omega,k) &=& S_{\psi}(\omega,k) + \sigma^2 \int dq \Biggl[
i D_F(\omega,q) \nonumber \\
&-& 4i\frac{\omega^2}{u_0^2}D_F(\omega,k)
D_F(\omega,q) \Biggr] h_{\Lambda}(k - q).
\label{p51}
\end{eqnarray}

Note that the induced interactions are different for the $\delta
\rho$ field from those for the $\psi$ field. This is due to the wave
equation for $\delta \rho$ is not of the form of Eq.~(\ref{eq-psi}),
since on replacing $\psi(t,z)  = u^2(z) \delta \rho(t,z)$ in this
equation there are extra spatial derivatives of $u(z)$. Therefore,
although similar, the model here is different from the ones studied
in condensed matter physics~\cite{gurarie}. As before, the colored
noise acts as a regulator of the loop integrals. Of course, there is
a natural cut-off in the high frequency regime, not related to
$\Lambda$, in the situation where the wavelength of the acoustic
phonon is of the order of the interatomic separation $l_{i}$, i.e.,
$\omega^{-1}\approx\,l_{i}$. For $\omega^{-1}<\,l_{i}$ the field
theoretical description of the acoustic excitation is no longer
valid.

In summary, we have constructed a condensed matter analog model for
fluctuations of the geometry of spacetime due to quantum gravity
effects. The model leads to the interpretation that such classical
background fluctuations induce effective interactions on free
fields. We expect that such sound cone fluctuations can be tested in
suspensions like colloids since excitations of acoustic modes in
these environments are described by random wave equations as the one
we discussed in this paper~\cite{ishimaru}. A possible course of
action is to remember the reference~\cite{gurarie}, where it is
shown that a non-linear dispersion relation generated by a random
wave equation induces peaks in the density of states. Another
direction is the analysis of localization of acoustic waves. Such
issue is related to the problem of how sound cone fluctuations in
our analog model can change the variation of flight of pulses. The
study of wave localization using diagrammatic perturbation method
was presented in the
Refs.~\cite{kirkpatrick,maynard,tamura1,tamura2}. It is known that
in one dimension any disorder is strong enough to induce exponential
localization of eigenmodes. On the other hand, a random fluid with a
supersonic acoustic flow is also an analog model that opens the
possibility to discuss the effect of the fluctuation of the geometry
in the Hawking radiation. Specific examples studying Anderson
localization in the analog model of the present paper as well as the
implementation of randomness in an acoustic black hole will be
reported elsewhere.

\newpage

\section{Acknowlegements}

N. F. Svaiter would like to acknowledge the hospitality of the
Instituto de F\'{\i}sica Te\'{o}rica, Universidade Estadual
Paulista, where part of this research was carried out. We would like
to thank L. H. Ford, S. E. Perez Bergliaffa and E. Arias Chinga for
useful discussions. This paper was supported in part by CNPq and
FAPESP (Brazilian agencies).


\begin{thebibliography}{99}
\bibitem{unruh1} W. G. Unruh, Phys. Rev. Lett. {\bf 46}, 1351
(1981).
\bibitem{hawking} S. W. Hawking, Comm. Math. Phys. {\bf 43}, 199
(1975).
\bibitem{mario} {\em{Artificial Black Holes}}, M. Novello, M.
Visser and G. Volovick (Editors), World Scientific, Singapure
(2002).
\bibitem{Barcelo:2005fc}
  C.~Barcelo, S.~Liberati and M.~Visser,
  Living Rev.\ Rel.\  {\bf 8}, 12 (2005)
.
%
%
\bibitem{arte} D. Arteaga, R. Parentani and H. Verdaguer,
Phys. Rev. {\bf D70}, 04409 (2004).
\bibitem{qg1} S. Weinfurtner, S. Liberati and M. Visser, J. Phys.
{\bf A39}, 6807 (2006).
\bibitem{ford11} L. H. Ford, Phys. Rev. {\bf D51}, 1692 (1995);
L. H. Ford and N. F. Svaiter, Phys. Rev. {\bf D56}, 2226 (1997); L.
H. Ford and N. F. Svaiter, Phys. Rev. {\bf D54}, 2640 (1996); H. Yu,
N. F. Svaiter and L. H. Ford, Phys. Rev. {\bf D80}, 124019 (2009).
\bibitem{hu1} B. L. Hu and K. Shiokawa, Phys. Rev. {\bf D57}, 3474
(1998).
\bibitem{principal} B. L. Hu and E. Verdaguer, Living Rev.
Relativity {\bf 7}, 3 (2004).
\bibitem{ford1} L. H. Ford and N. F. Svaiter, Phys. Rev. Lett.
{\bf 102}, 030602 (2009).
\bibitem{lifshitz} I. E. Dzyaloshinskii, E. M. Lifshitz and L. P.
Pitaevski, Adv. Phys. {\bf 10}, 165 (1961).
\bibitem{ford2} L. H. Ford and N. F. Svaiter, Phys. Rev. {\bf
D80}, 065034 (2009).
\bibitem{fluid} L. D. Landau and E. M. Lifshitz, {\em{Fluid
Mechanics}}, Elsevier (1987).
\bibitem{gurarie} V. Gurarie and V. Altland, Phys. Rev. Lett. {\bf
94}, 245502 (2005); S. John, H. Sompolinsky and M. J. Stephen, Phys.
Rev. {\bf B27}, 5592 (1983); S. John and M. J. Stephen, Phys. Rev.
{\bf B28}, 6358 (1983).
\bibitem{gorkov} A. A. Abrikosov, L. P. Gorkov and I. E.
Dzyaloshinski, {\em{Methods of Quantum Field Theory in Statistical
Physics}}, Dover Publication Inc. New York (1975).
\bibitem{papa} J. Pierre Fouque, J. Garnier, G.
Papanicolaou and K. Solna, {\em {Wave Propagation and Time Reversal
in Randomly Layered Media}}, Springer Science and Bussiness Media,
LLC (2007).
\bibitem{lodahl} P. Lodahl, A. P. Mosk and A. Lagendijk,
Phys. Rev. Lett {\bf 95}, 173901
(2005).
\bibitem{skipetrov} S. E. Skipetrov, Phys. Rev. A {\bf 75}, 053808 (2007).
%
%
\bibitem{ishimaru} A. Ishimaru, {\em {Wave Propagation and Scattering in Random Media
}}, IEEE, New York (1997).
\bibitem{kirkpatrick} T. R. Kirkpatrick, Phys.
Rev. {\bf B31}, 5746 (1985).
\bibitem{maynard} E. Akkermans and R. Maynard, Phys.
Rev. {\bf B32}, 7850 (1985).
\bibitem{tamura1} N. Nishiguchi, S. Tamura and F. Nori, Phys. Rev. {\bf B41},
7941 (1990).
\bibitem{tamura2} S. Tamura and F. Nori, Phys. Rev. {\bf B48},
2515 (1993).

\end{thebibliography}
\end{document}